\documentclass[aps,prl,twocolumn,groupedaddress,showpacs]{revtex4}

\usepackage{graphicx}

\draft

\begin{document}

\title{
Nondissipative Spin Hall Effect via Quantized Edge Transport}

\author{L. Sheng$^{1}$, D. N. Sheng$^2$, C. S. Ting$^1$ and F. D. M.
Haldane$^3$}

\address{
$^1$Department of Physics and Texas Center for Superconductivity,
University of Houston, Houston, Texas 77204\\
$^2$Department of Physics and Astronomy, California State
University, Northridge, California 91330\\
$^3$Department of Physics, Princeton University, Princeton, NJ
08544}

\begin{abstract}

The spin Hall effect in a two-dimensional electron system on
honeycomb lattice with both intrinsic and Rashba spin-orbit
couplings is studied numerically. Integer quantized spin Hall
conductance is obtained  at zero Rashba coupling limit when
electron Fermi energy lies in the energy gap created by the
intrinsic spin-orbit coupling, in agreement with recent
theoretical prediction. While nonzero Rashba coupling destroys
electron spin conservation, the spin Hall conductance is found to
remain near the quantized value, being insensitive to disorder
scattering, until the energy gap collapses with increasing the
Rashba coupling. We further show that the charge transport through
counterpropagating spin-polarized edge channels is well quantized,
which is associated with a topological invariant of the system.
\end{abstract}

\mbox{}\\

 \maketitle





The proposals of intrinsic spin Hall effect (SHE) in a Luttinger
spin-orbit (SO) coupled three-dimensional $p$-doped
semiconductor~\cite{t1} and in a Rashba SO coupled two-dimensional
electron system (2DES)~\cite{t2} have stimulated many subsequent
research activities~\cite{t4,t5,t6,t7,t8,t9,
t11,s11,t12,t13,t14,t15,t18,t19,t17,t21}. The SHE may potentially
provide a purely electrical means to manipulate electron spins
without use of ferromagnetic materials or a magnetic field. The
SHE in these systems is dissipative because of nonzero
longitudinal conductance~\cite{t11} and exhibits nonuniversal
behavior in the presence of
disorder~\cite{t6,t7,s11,t12,t13,t14,t15,t18,t19,t21}, which is
naturally distinct from the conventional integer quantum Hall
effect (IQHE). In particular, it is found~\cite{t6,t7,t19} that
the bulk SHE in two-dimensional Rashba model may be destroyed by
any weak disorder in infinite samples. It is of both fundamental
and practical interest to search for nondissipative SHE with
universal properties similar to the IQHE, in light of that IQHE
can exist in nature in the absence of magnetic field, as first
predicted by Haldane~\cite{t23}.

A class of band insulators with SO coupling were suggested as
possible candidates for nondissipative SHE~\cite{t11}.
Interestingly, Kane and Mele proposed~\cite{t22} that the
intrinsic SO coupling in single-layer graphene films may give rise
to an integer quantized SHE (IQSHE). The intrinsic SO coupling
conserves electron spin $s_z$. The independent subsystems of two
spin directions $\sigma=\uparrow$ and $\downarrow$ are each
equivalent to Haldene's spinless IQHE model~\cite{t23} on
honeycomb lattice without magnetic field. They contribute
quantized Hall conductances $e^2/h$ and $-e^2/h$, respectively,
when the electron Fermi energy lies inside the energy gap created
by the SO coupling. While the charge Hall conductances cancel out,
the total spin Hall conductance (SHC) is quantized to
$\sigma_{sH}=2$ in units of $e/4\pi$. We recall that each
subsystem can be classified by an integer Chern number~\cite{t23},
which equals the Hall conductance of the subsystem in units of
$e^2/h$, and is conserved without spin-mixing interactions. Upon
coupling the two subsystems, only the total Chern number (as a
well-known topological invariant) is conserved , which is
trivially zero as the total Hall conductance vanishes. Therefore,
the conservation of electron $s_z$ appears to be important to the
IQSHE. It is unclear whether the topological SHE could survive if
electron $s_z$ conservation is destroyed, e.g., by the Rashba SO
coupling, which usually exists in a 2DES due to asymmetry in the
confining potential. Furthermore, disorder effect in the class of
insulating SHE systems has not been studied so far. It is highly
desirable to investigate these important issues.

In this Letter,  the SHE in the 2D honeycomb lattice  model
including the intrinsic and Rashba SO couplings is studied
numerically. By using the multi-probe Landauer-B\"{u}ttiker
formula, we show that the SHC remains nearly quantized in the
presence of finite Rashba coupling and disorder scattering until
the energy gap collapses. We further show that the  charge
transport through spin-polarized edge channels is well quantized
even for nonzero Rashba coupling, which accounts for the
robustness of the SHE. The SHC in samples with close boundary
conditions is also calculated by using the Kubo formula, whereby
the SHE is shown to be a stable bulk effect instead of a boundary
effect. Our work provides the first numerical demonstration of
robust nondissipative SHE in a spin nonconservative 2DES in the
presence of disorder. The nontrivial topological origin of this
nondissipative transport regime is also discussed.

The Hamiltonian for a 2DES on a honeycomb lattice can be written
as~\cite{t23,t22,t24}
\begin{eqnarray}
H&=&-t\sum\limits_{\langle ij\rangle}c_{i}^\dagger
c_{j}+\frac{2i}{\sqrt{3}}V_{\mbox{\tiny SO}}
\sum\limits_{\langle\langle ij\rangle\rangle}c_{i}^\dagger
{\mbox{\boldmath$\sigma$}}\cdot({\bf d}_{kj}\times
{\bf d}_{ik})c_{j}\nonumber\\
&+&iV_{\mbox{\tiny R}}\sum\limits_{\langle ij\rangle}c_{i}^\dagger
\hat{\bf e}_z\cdot({\mbox{\boldmath$\sigma$}} \times{\bf
d}_{ij})c_{j}+\sum\limits_{i}\epsilon_ic_{i}^\dagger c_{i}\
,\label{HAMIL}
\end{eqnarray}
where $c_{i}^\dagger=(c_{i\uparrow}^\dagger,
c_{i\downarrow}^\dagger)$ are electron creation operators, and
${\mbox{\boldmath$\sigma$}}$ are the Pauli matrices. The first
term is the usual nearest neighbor hopping term. The second term
is the intrinsic SO coupling allowed by the symmetries of the
honeycomb lattice~\cite{t23,t22,t24} with $i$ and $j$ as two next
nearest neighbor sites, where $k$ is the only common nearest
neighbor of $i$ and $j$, and ${\bf d}_{ik}$ is a vector pointing
from $k$ to $i$. The third term is the Rashba SO coupling with
$\hat{\bf e}_z$ a unit vector in the $z$ direction. The last term
describes nonmagnetic disorder, where $\epsilon_i$ is a random
on-site potential uniformly distributed in the interval $[-W/2,
W/2]$. The distance between nearest neighbor sites is taken to be
unity. We mention that honeycomb lattice may be realized in other
microstructures as well as in single-layer graphene
films~\cite{tt24}. For example, in a triangular antidot lattice
created at a semiconductor heterointerface by using artificial
periodic repulsive potentials, the electrons can be restricted
into the region of a honeycomb sublattice~\cite{antidot}. If we
switch off the Rashba coupling by setting $V_{\mbox{\tiny R}}=0$,
Eq.\ (\ref {HAMIL}) reduces to a two-component Haldane's
model~\cite{t23}, which is expected to display $\sigma_{sH}=2$
IQSHE~\cite{t22}.

\begin{figure}
\includegraphics[width=2.5in]{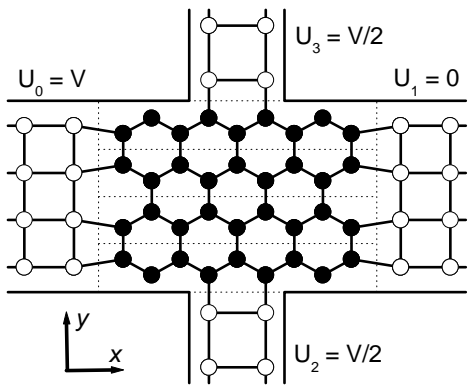}
\caption{The four-probe spin Hall bar setup used for calculating
the SHC. Filled circles represent the sites in the sample, and
opened circles stand for the sites in the leads. $U_l$ is the
electrical voltage in lead $l$.}
\end{figure}
We consider a four-probe spin Hall setup as illustrated in Fig.\
1, where a rectangular honeycomb lattice sample is connected with
four ideal semi-infinite leads. The actual system used in our
calculations will have the same aspect ratios as Fig.\ 1 but
enlarged sizes. To specify the system size, the sample is divided
into $L_y$ horizontal chains with $L_x$ sites in each chain, as
indicated by the dotted lines in Fig.\ 1. The total number of
sites in the sample is denoted as $N=L_x\times L_y$. For
simplicity, the leads are assumed to have a square lattice
structure with only nearest neighbor hopping $t$. The spin
currents are well defined in the leads, where no SO interactions
exist. The SHC is given by twice the ratio of the spin current in
lead $3$ to the voltage drop $V$ between leads $0$ and $1$. Here,
a factor $2$ is used to properly eliminate the effect of the
contact resistances between the leads and the edge channels in the
four-probe setup~\cite{t22}. The linear SHC is calculated exactly
by using the multi-probe Landauer-B\"{u}ttiker
formula~\cite{t13,t14,t15,t25}
\begin{equation}
\sigma_{sH}=\frac{e}{4\pi}\sum\limits_{\sigma\sigma'}\sigma
\left(T^{\sigma\sigma'}_{30}-T^{\sigma\sigma'}_{31}\right)\
,\label{SHC4Term}
\end{equation}
where $T_{ll'}^{\sigma\sigma'}$ is the spin-resolved electron
transmission coefficient from spin $\sigma'$ channels in lead $l'$
to spin $\sigma$ channels in lead $l$ at the Fermi energy $E$.

\begin{figure}
\includegraphics[width=3.0in]{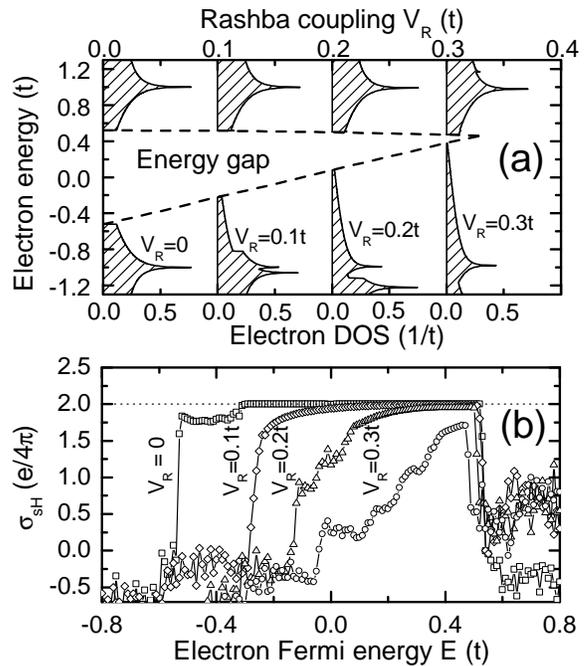}
\caption{(a) The curves with filled areas are electron DOS (bottom
axis) in a clean bulk sample versus electron energy (left axis)
for four different Rashba coupling strengths $V_{\mbox{\tiny R}}$.
The dashed lines represent the two edges of the energy gap as
functions of the Rashba coupling strength $V_{\mbox{\tiny R}}$
(top axis). (b) Four-probe SHC versus electron Fermi energy for
some $V_{\mbox{\tiny R}}$ calculated on a $N=129\times 64$ sample.
For both (a) and (b), $W=0$ and $V_{\mbox{\tiny SO}}=0.1t$.}
\end{figure}
In Fig.\ 2a, we show the electron density of states (DOS)
calculated in the momentum space for a large clean bulk sample.
The SHE within the energy gap is of our main interest. For a
sample at half filling, such as an undoped graphene film, the
presence of weak disorder can pin the electron Fermi energy inside
the gap. We note that the actual electron DOS of the sample in the
four-probe setup may be slightly modified from that shown in Fig.\
2a because of the connection with the leads. In Fig.\ 2b, the
calculated SHC is shown as a function of electron Fermi energy $E$
for sample size $N=129\times 64$ and several different strengths
of the Rashba coupling $V_{\mbox{\tiny R}}$. At $V_{\mbox{\tiny
R}}=0$, the energy gap in the DOS for a clean bulk sample is
between $-0.52t$ to $0.52t$. As expected, the SHC in Fig.\ 2b is
well quantized to integer $2$ in the main region of the gap. As
$V_{\mbox{\tiny R}}$ increases to $0.1t$, the gap shrinks to
$-0.23t$ to $0.51t$. The SHC within the gap deviates from the
quantized value slightly, showing very small fluctuations with
$E$, in contrast to the strong fluctuations outside the gap. The
same feature is observed for $V_{\mbox{\tiny R}}=0.2t$ and $0.3t$.

\begin{figure}
\includegraphics[width=3.0in]{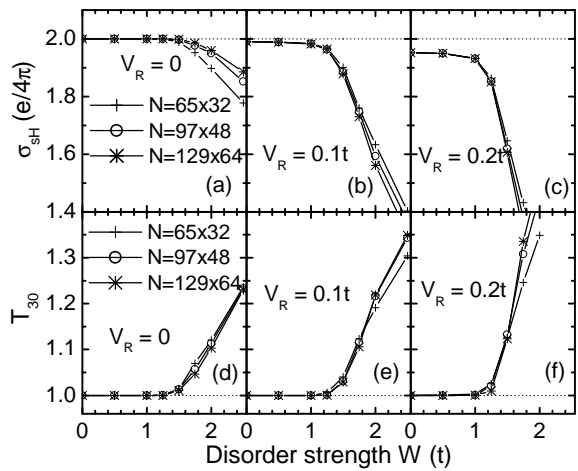}
\caption{Four-probe SHC $\sigma_{sH}$ and total transmission
coefficient $T_{30}$ for $V_{\mbox{\tiny SO}}=0.1t$, $E=0.4t$ and
three different sample sizes. (a) to (c): $\sigma_{sH}$ versus
disorder strength $W$ for $V_{\mbox{\tiny R}}=0$, $0.1t$ and
$0.2t$, respectively. (d) to (f): corresponding total transmission
coefficient $T_{30}$ from lead $0$ to lead $3$ versus $W$. Here,
disorder average is performed over 1000 random configurations.}
\end{figure}
The effect of disorder is studied by fixing electron Fermi energy
at $E=0.4t$. For $V_{\mbox{\tiny R}}<0.3t$ considered below,
$E=0.4t$ is always inside the energy gap of a pure bulk system.
The SHC calculated for three different $V_{\mbox{\tiny R}}$ are
plotted in Fig.\ 3a-3c, respectively, as functions of $W$. The
results for three different sample sizes ($N=65\times 32$,
$97\times 48$ and $129\times 64$) are shown together for
comparison. In Fig.\ 3d-3f, the corresponding total transmission
coefficient $T_{30}=\sum_{\sigma\sigma'}T^{\sigma\sigma'}_{30}$ is
plotted. All $T_{ll'}$ for neighboring leads $l$ and $l'$ are
equal after disorder average by symmetry. They characterize the
charge transport between the leads.

Let us first consider the case of zero Rashba coupling
$V_{\mbox{\tiny R}}=0$, where the SHC is integer quantized at zero
disorder. According to Fig.\ 3a, the IQSHE persists for a range of
disorder strength $0\leq W\lesssim 1.2t$. In the same range,
$T_{30}$ in Fig.\ 3d is also quantized to $1$. This is not
surprising because at $V_{\mbox{\tiny R}}=0$ the subsystems of the
two spin directions are two decoupled IQHE systems. Our result is
consistent with electron transport through fully spin-polarized
edge channels with spin-dependent chirality. In the IQSHE regime,
our calculation yields
$T^{\uparrow\uparrow}_{30}=T^{\uparrow\uparrow}_{13}
=T^{\uparrow\uparrow}_{21}=T^{\uparrow\uparrow}_{02}=1$,
representing a left-moving edge mode in $\sigma=\uparrow$
subsystem, and
$T^{\downarrow\downarrow}_{03}=T^{\downarrow\downarrow}_{31}
=T^{\downarrow\downarrow}_{12}=T^{\downarrow\downarrow}_{20}=1$,
corresponding to a right-moving edge mode in $\sigma=\downarrow$
subsystem. All the other spin-resolved transmission coefficients
vanish. Strong disorder $W\gtrsim 1.2t$ destroys the quantizations
of the SHC and transmission coefficients. On the strong disorder
side, $T_{30}$ increases rather than decreases with increasing
$W$, which signals the collapse of the bulk mobility gap. It is
verified but not shown here that, with further increasing $W$, all
the transmission coefficients eventually decrease to zero because
of electron localization.

Next we look at how the SHC and charge transport evolve with
disorder at nonzero Rashba coupling. Remarkably, we see from Fig.\
3e and 3f that $T_{30}$ is still well quantized within a
relatively small range of $W$, indicating that the edge modes
remain robust. $T_{30}(=T_{13}=T_{21}=T_{02})=1$ relates to the
left-moving mode, and $T_{03}(=T_{31} =T_{12}=T_{20})=1$ relates
to the right-moving mode. However, the spin-resolved transmission
coefficients are no longer quantized. For example, at
$V_{\mbox{\tiny R}}=0.2t$, $W=0.5t$ and $N=129\times 64$, we have
$T_{30}^{\uparrow\uparrow}=0.960$,
$T_{30}^{\uparrow\downarrow}=0.028$,
$T_{30}^{\downarrow\downarrow}=0.000$ and
$T_{30}^{\downarrow\uparrow}=0.012$, suggesting that the edge
modes become partially spin-polarized. Nonetheless, as long as the
charge transport is quantized, the SHC stays near the quantized
value ($\sigma_{sH}=1.95$ for the above parameters), and is robust
as it is insensitive to disorder $W$ and independent of sample
size $N$, as seen from Fig.\ 3b and 3c. With further increasing
$W$, $T_{30}$ deviates from the quantized value, and the SHC also
decreases rapidly, the system undergoing a transition to a
dissipative transport regime.

We have observed that in the presence of not too strong Rashba
coupling and disorder, the SHE remains robust and nearly integer
quantized.  Similarly to the IQHE, while the effective
current-carrying states are edge states in open-boundary systems,
the nearly quantized SHE is essentially a stable bulk effect
insensitive to boundary conditions or local Hamiltonian at the
boundary. It is of interest to demonstrate this point directly,
especially, in view of the sensitivity to boundary conditions of
the SHE in other metallic models~\cite{t13,t14,t15,t21}. We
consider a square sample without leads. Periodic boundary
conditions are imposed in both the $x$ and $y$ directions. This
close-boundary system has translational invariance in the absence
of disorder. The Kubo formula~\cite{t2} is conveniently used to
calculate the SHC $\sigma_{sH}$ by exact diagonalization of the
system Hamiltonian~\cite{t19}. Standard spin current
operator~\cite{t2} $J_{ys}^z=(s_zv_y+v_ys_z)/2$ is adopted, where
$v_y$ is the electron velocity operator in the $y$ direction.

\begin{figure}
\includegraphics[width=3in]{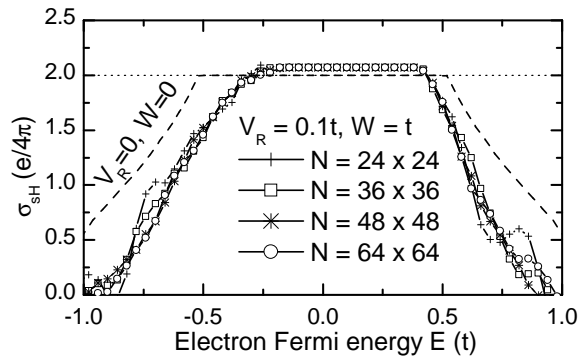}
\caption{SHC calculated from the Kubo formula as a function of
electron Fermi energy $E$ for $V_{\mbox{\tiny SO}}=0.1t$. The
dashed line is $\sigma_{sH}$ in a clean sample ($W=0$) with
$V_{\mbox{\tiny R}}=0$ and $N=64\times 64$. The lines with symbols
are $\sigma_{sH}$ at  $W=t$ and $V_{\mbox{\tiny R}}=0.1t$ for four
different sample sizes. Here, $\sigma_{sH}$ is averaged over 200
disorder configurations for $N=64\times 64$ and 1000 disorder
configurations for the smaller samples.}
\end{figure}
In Fig.\ 4, the dashed line is the calculated $\sigma_{sH}$ for
the ideal case of zero disorder $W=0$ and Rashba coupling
$V_{\mbox{\tiny R}}=0$ for system size $N=64\times 64$.
$\sigma_{sH}$ is well quantized to $2$ in units of $e/4\pi$ within
the energy gap ($-0.52t$ to $0.52t$). The lines with symbols are
$\sigma_{sH}$ at nonzero $W=t$ and $V_{\mbox{\tiny R}}=0.1t$ for
four different sample sizes from $N=24\times 24$ to $64\times 64$.
In comparison with the ideal case, the SHC in the gap ($-0.23t$ to
$0.51t$ for a pure sample) is not well quantized because of
nonzero $V_{\mbox{\tiny R}}$. However, the SHC is very close to
the quantized value even in the presence of disorder $W=t$. The
four lines collapse in the middle region, an indication that the
nearly quantized SHC does not change with increasing sample size,
and is thus expected to persist in the thermodynamic limit. We
have seen that the SHC in this close-boundary system behaves
similarly to that in the four-probe setup, provided that the
electron Fermi energy lies inside the energy gap. A nonessential
discrepancy is observed that the nearly quantized SHC at
$V_{\mbox{\tiny R}}>0$ for the former system is a little greater
than $2$ ($\sigma_{sH}=2.07$ in Fig.\ 4) and that for the latter
system is smaller than $2$. We believe that this discrepancy is
caused by the definition of bulk spin current $J_{ys}^z$, which is
not conservative~\cite{t7} and hence is not completely equivalent
to the spin current measured in leads.

The robust SHE and the quantized charge transfer through edge
channels are associated with the nontrivial topological properties
of the honeycomb lattice model with SO couplings. In the absence
of the Rashba coupling, we have two decoupled subsystems of spin
$\sigma=\uparrow$ and $\downarrow$, and each exhibits an integer
quantized Hall conductance, $e^2/h$ and $-e^2/h$. Each subsystem
can be classified by an integer Chern number, with $C_{\uparrow} =
-C_{\downarrow}=1$. In an open system, there will be two decoupled
chiral edge modes moving in opposite directions along the
boundary. This picture is substantially altered by the Rashba
coupling when the mirror-plane symmetry is destroyed. There are no
longer two ``types'' of electrons due to spin-mixing effect of the
Rashba term. However, while the total Chern number vanishes, the
opposite nonzero Chern numbers in the coupled system can not
annihilate each other, as a consequence of ``parity anomaly'' in
the decoupled limit~\cite{t23}. They coexist and lead to a new
topological invariant, which manifests as a pair of edge modes
with partial spin polarizations, as indicated by the numerical
results. These edge modes are robust in the presence of disorder,
until the energy gap collapses, where the low-energy states merge
with their high-energy parity partners~\cite{t23}. It is worth
stressing that the SHC itself is not a topological invariant,
which decreases continuously with increasing the strength of the
Rashba coupling as the edge states become less spin-polarized.
Mathematical description of the new topological invariant will be
reported elsewhere.

\textbf{Note added:} As we are finishing this paper, it is
interesting to notice that in a couple of recent
preprints~\cite{tt21}, different models for IQSHE are proposed and
studied for pure systems. Our paper represents the first numerical
work on the characterization of the SHE in this class of models in
the presence of random disorder and coupling between different
topological subsystems.

\textbf{Acknowledgment:} We thank C. L. Kane and Z. Y. Weng for
stimulating discussions. This work is supported by ACS-PRF
41752-AC10, Research Corporation Fund CC5643, the NSF
grant/DMR-0307170 (DNS), a grant from the Robert A. Welch
Foundation (CST), and NSF (under MRSEC grant/DMR-0213706) at the
Princeton Center for Complex Materials (FDMH).

\end{document}